\documentclass[journal]{IEEEtran}

\DeclareUnicodeCharacter{2212}{-}
\usepackage{xcolor,soul,framed} %,caption

\colorlet{shadecolor}{yellow}
\usepackage[pdftex]{graphicx}
\graphicspath{{../pdf/}{../jpeg/}}
\DeclareGraphicsExtensions{.pdf,.jpeg,.png}

\usepackage[cmex10]{amsmath}
\usepackage{array}
\usepackage{mdwmath}
\usepackage{mdwtab}
\usepackage{eqparbox}
\usepackage{url}

\begin{document}
%\Large -> Fontsize 12
%\title{\Large Transition-state-theory based interpretation of Landau double well potential for ferroelectrics: No “Quasi-Static Negative capacitance” \vspace{-0.4 \baselineskip}}
%\LARGE -> Fontsize 14

%\title{\LARGE Transition-state-theory based interpretation of Landau double well potential for ferroelectrics: No “Quasi-Static Negative capacitance” \vspace{-0.4 \baselineskip}}

%\huge -> Fontsize 14

\title{\huge Transition-state-theory-based interpretation of Landau double well potential for ferroelectrics \vspace{-0.4 \baselineskip}}

\author{
\IEEEauthorblockN{Md~Nur~K.~Alam \IEEEauthorrefmark{2}\IEEEauthorrefmark{4},~S.~Clima \IEEEauthorrefmark{1},~ B.~Kaczer \IEEEauthorrefmark{1},~ Ph.~Roussel \IEEEauthorrefmark{1},~ B.~Truijen 
\IEEEauthorrefmark{1},~ L.~\AA.~Ragnarsson 
\IEEEauthorrefmark{1},~ N.~Horiguchi \IEEEauthorrefmark{1}, ~M.~Heyns \IEEEauthorrefmark{3}, and J.~Van~Houdt \IEEEauthorrefmark{1}\IEEEauthorrefmark{3}}

\IEEEauthorblockA{
    \IEEEauthorblockA{\IEEEauthorrefmark{1}IMEC, Belgium}
    \IEEEauthorrefmark{2} Khulna University of Engineering \&\ Technology, Bangladesh}
    \IEEEauthorblockA{\IEEEauthorrefmark{3}KU Leuven, Belgium}
    \IEEEauthorblockA{\IEEEauthorrefmark{4}nur.kutubul.alam@eee.kuet.ac.bd}
}

\IEEEaftertitletext{\vspace{-2.5\baselineskip}}

\maketitle

% === ABSTRACT ====================================================================
% =================================================================================

\begin{abstract}
%\boldmath
Existence of quasi-static negative capacitance (QSNC) was proposed from an interpretation of the widely accepted Landau model of ferroelectrics. However, many works showed not to support the QSNC theory, making it controversial. In this letter we show the Landau model when used together with transition-state-theory, can connect various models including first-principles, Landau, Preisach and nucleation limited switching while it does not predict the existence of QSNC. 
\end{abstract}

% === KEYWORDS ====================================================================
% =================================================================================

% \fi
\IEEEpeerreviewmaketitle

% ====================================================================
% ====================================================================
% ====================================================================

% === I. INTRODUCTION =============================================================
% =================================================================================
\section{Introduction}

\IEEEPARstart{S}{teep}-subthreshold-slope in the transfer characteristics of MOSFETs is needed to keep on scaling the supply voltage ($V_{DD}$) that would in turn reduce power dissipation without sacrificing performance. In 2008 \cite{Salahuddin_first_paper} it was proposed from an extended interpretation of the Landau model that QSNC exists in ferroelectric (FE) materials that can provide the required steep slope operation of MOSFETs. 
Many mathematical models are available for describing ferroelectricity. Physics-based first principles models like DFT gives clear insight to the working of FE at a large computational cost. On the other hand, the least expensive phenomenological models (e.g., Preisach, KAI, NLS \cite{Alam_HZO_arbitrary_exciation}) do not pay serious attention to the underlying physics. Landau model was originally designed for describing phase transition phenomena using a double well thermodynamic potential. Minimization of that potential gives an S-shaped polarization-field relation, largely referred to as the “S-curve”. QSNC theory states a FE is stabilizable on the negative slope region of the S-curve by applying a charge boundary condition (CBC) by stacking the FE with a dielectric (DE) and/or with a semiconductor channel of a MOSFET \cite{Salahuddin_first_paper}. 
The QSNC theory was not universally accepted by the scientific community. Experimental works that claimed to capture any signature of QSNC \cite{Khan_NC_in_FE_cap_Nature} or the S-curve \cite{Hoffmann_S_curve} have been shown to be a misinterpretation of measurement data, independently by D. A. Antoniadis from MIT \cite{Alamo_Switching_dynamics}, T. P. Ma from Yale \cite{TP_Ma_Critical_exam_QSNC}, J. A. Kittl from Samsung \cite{Kittl_Hoffmann_debunked} and many others. We also showed the same by direct application of CBC \cite{Alam_Nature}. It is yet to be shown whether prediction of the QSNC is intrinsic to the Landau model or not.
In this article we show that the double well potential comes from the physics-based model. A phenomenological adjustment translates it to the Landau model with a better quantitative predictability. Finally, we describe the polarization switching dynamics using transition state theory taking the Landau potential as two-state system. Such a model bridges all the model types, correctly predicts experimental outcomes and does not predict the existence of QSNC. We keep the discussion particularly for $HfO_2$ based ferroelectric as it is technologically most relevant, however the outcome of the work is general for any FE system.

% === II. Physics based atomistic model to Landau model ========================
% =================================================================================
%\section{Physics based atomistic model to Landau model}
\section{Atomistic model to Landau model}
\begin{figure}
   \begin{center}
   \includegraphics[width=3.4in]{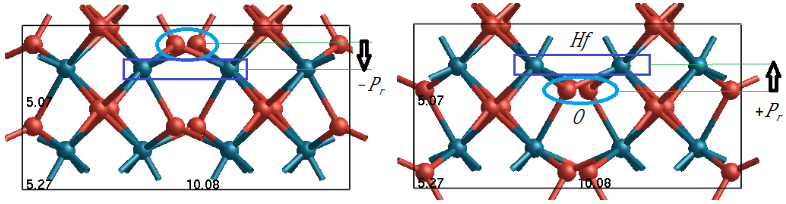}\\
   \caption{Two possible minimum energy atomic configurations of orthorhombic $HfO_2$ with up and down polarization.}\label{Atomic_structure_with_P}
   \end{center}
 \end{figure}

Two minimum energy configurations of orthorhombic $HfO_2$ is shown in Fig. \ref{Atomic_structure_with_P}, 
Orthrhombic $HfO_2$ has two possible minimum energy atomic configuration, where the displacement between $Hf$ and $O$ atoms in each of them give rise to polarization $\pm P$. Switching between $+P$ and $-P$ or vice-versa involves transformation between two structures \cite{Sergiu_IEDM_18}. We calculate the minimum energy path needed for the process to occur using a Nudged Elastic Band (NEB) approach with the climbing image algorithm \cite{CI_algorithm}. During switching, energy and $P$ of images (snapshots of the atomic positions during switching) are computed by density functional theory (DFT) with a 2×2×2 Monkhorst–Pack k-point mesh using mixed (Gaussian and plane wave) basis within local density approximation (LDA) and Berry phase approach, respectively. 

\begin{figure}
  \begin{center}
  \includegraphics[width=3.4in]{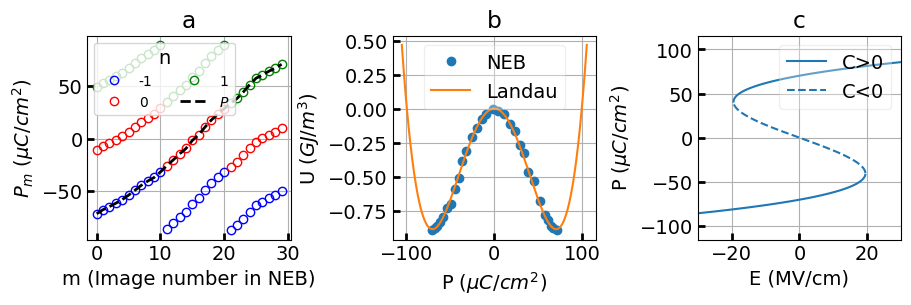}
 % \vspace{-15pt}
  \caption{a) Polarization in the $m^{th}$ image of the NEB simulation, given by equation 1. Different parallel branches are created due to the uncertainty imposed by the quantum of polarization, b) Energy profile as a function of $P$ (from Fig. \ref{Landau_profile_from_NEB}a) fitted with Landau model, c) Minimization of the Landau potential gives the S-curve }\label{Landau_profile_from_NEB}
  \end{center}
\end{figure}

Fig. \ref{Landau_profile_from_NEB}a shows multiple branches of continuously varying $P_m$ ($P$ of $m^{th}$ image). It is attributed to the uncertainty in $P$ computation caused by crystal's periodicity, given by \cite{Quantum_of_P}

%Continuous motion of $O$ atoms must give a continuous change of $P_m$ ($P$ of $m^{th}$ image). Fig. \ref{Landau_profile_from_NEB}a shows multiple branches of continuously varying $P_m$. It is attributed to the uncertainty in $P$ caused by crystal's periodicity, given by \cite{Quantum_of_P} $P_m$ ($P$ of $m^{th}$ image) shows multiple brunches

%Fig. \ref{Atomic_structure_with_P} shows two possible minimum energy configurations of the orthorhombic $HfO_2$ crystal obtained by relaxing the atomic positions of $Hf$ and $O$. The displaced oxygen atoms from a plane of $Hf$ atoms give rise to a polarization $\pm{P}$. Switching the polarization direction from $+P$ to $-P$ or vice-versa involves transformation between these two structures under the influence of external field \cite{Sergiu_IEDM_18}. We calculate the minimum energy path needed for the process to occur using a Nudged Elastic Band (NEB) approach with the climbing image algorithm \cite{CI_algorithm} in CP2K package. Energy is computed by density functional theory (DFT) with a 2×2×2 Monkhorst–Pack k-point mesh using mixed (Gaussian and plane wave) basis, within local density approximation (LDA). The images are equivalent to snapshots of the atomic positions during switching. The $P$  in each of the images has been calculated using the Berry phase approach (valid for bulk structure). Computed $P$ of the $m^{th}$ image, $P_m$, is periodic due to the periodicity of the crystal and is given by \cite{Quantum_of_P}-

\begin{equation}\label{Pm_periodic_equation}
P_m = P_m +nQ_n
\end{equation}

where $Q_n$ is “quantum of polarization”, and $n$ is an integer. Image-1 and 29 must give remnant polarization, $-P_r$ and $+P_r$ respectively, as they correspond to the two minimum energy atomic configurations (Fig. \ref{Atomic_structure_with_P}). Therefore, the symmetric branch in Fig. \ref{Landau_profile_from_NEB}a gives $P$.

%where $Q_n$ is “quantum of polarization”, and $n$ is an integer. Fig. \ref{Landau_profile_from_NEB}a   shows a discontinuous jump between the $10^{th}$ and the $11^{th}$ images and between $20^{th}$ and the $21^{st}$ images, for a given value of $n$. It happens due to the uncertainty in polarization computation by an amount  given by eq. \ref{Pm_periodic_equation}. But motion of $O$ atoms is continuous, and so the change of $P_m$. Also, image-0 and image-29 being two minimum energy configurations they give the remnant polarization, $-P_r$ and $+P_r$, respectively. Therefore, the symmetric branch in Fig. \ref{Landau_profile_from_NEB}a gives $P$. 

% =======
% FIG. 01
% =======

%Fig. \ref{Landau_profile_from_NEB}b shows computed energy (at temperature T=0K) of the atomic structures (images) as a function of $P$, where two minimum energy states are separated by an energy barrier. The vibrational eigen mode of the minimum energy images and the one corresponding to the top of the barrier (TOB) are found to be 112.532 $cm^{−1}$ and 434.269i $cm^{−1}$, respectively. The imaginary vibrational mode proves the TOB (with $P=0$ in Fig. \ref{Landau_profile_from_NEB}b) is a saddle point. Remnant polarization $P_r$ is found to be 71 $\mu C/cm^2$. Using a crude approximation \cite{Sergiu_dielectric_constant_near_switching} that the dielectric response is constant $\epsilon_{FE}=35$, the intrinsic coercive field $E_C$ from half depolarization field model \cite{Intrinsic_Ec} is found to be 11 $MV/cm$. In contrast, measurement from IMEC’s 10nm thick $HfO_2$ capacitor shows a $P_r$ and $E_C$ are 15 $\mu C/cm^2$ and 1.1 $MV/cm$, respectively.  This shows that it is difficult to predict  without considering the domain dynamics. But NEB simulation gives a good insight that FE $HfO_2$ minimizes the energy of the crystal in an atomic configuration.

Fig. \ref{Landau_profile_from_NEB}b shows computed energy (at temperature T=0K) as a function of $P$. Results obtained from NEB is fitted with Landau model (Fig. \ref{Landau_profile_from_NEB}b) that gives the energy as a function of $P$ as-

%The NEB data is fitted with Landau model (Fig. \ref{Landau_profile_from_NEB}b) that gives the energy as a function of $P$ as-

\begin{equation}\label{Landau_eq_for_Free_Energy}
U = \alpha P^2 + \beta P^4 - EP
\end{equation}

%Here $E$ is applied electric field, $\alpha$ and $\beta$ are Landau parameters. Their values are found to be 3.56207142×$10^9$ ($C^{−2}Jm$) and 3.56045779×$10^9$ ($C^{−4}Jm^5$) respectively. The fitting also gives the characteristic double-well potential. For a better quantitative description of $P_r$ and $E_C$, the $\alpha$ and $\beta$ parameters are phenomenologically adjusted. In doing so, the shape of Fig. \ref{Landau_profile_from_NEB}b and the underlying physics are kept unchanged while the value of $P$ in the abscissa is adjusted. The $P$-$E$ relation at steady-state is obtained by minimizing eq. \ref{Landau_eq_for_Free_Energy}, that gives-

Here $E$ is applied electric field, $\alpha$ and $\beta$ are Landau parameters. Minimization of $U$ with respect to $P$ gives $P-E$ relationship at steady-state as shown in Fig. \ref{Landau_profile_from_NEB}c, also referred to as “S-curve”. The broken line, that has a negative slope and therefore QSNC, represents the maximum in the middle of the Landau potential where the curvature of the profile is negative (Fig. \ref{Landau_profile_from_NEB}b).

%\begin{equation}\label{PE_from_Landau_minimization}
%E = 2 \alpha P + 4 \beta P^3 
%\end{equation}

%The  relation from eq. \ref{PE_from_Landau_minimization} is shown in Fig. \ref{Landau_profile_from_NEB}c. The sections in solid line represent the minima in Fig. \ref{Landau_profile_from_NEB}b. The broken line, that has a negative slope and therefore QSNC, represents the maximum in the middle of the Landau potential where the curvature of the profile is negative (Fig. \ref{Landau_profile_from_NEB}b).  

\section {Phenomenological/Statistical model}

%The NEB simulates a large $HfO_2$ system by applying a periodic boundary condition, where each of the unit cells is interacting with periodic images. Such interaction makes notable effect within a finite size of the system, known as the \textit{correlation length}. A large system always spontaneously breaks into many independent elementary regions or domains. We showed Preisach (a steady-state model) and NLS (transient model) can describe the statistical behavior of many domains of $HfO_2$ at a low computational cost \cite{Alam_HZO_arbitrary_exciation}. 

Small regions of ferroelectric can switch their $P$ independent of each others, called hysterons. Their statistical behavior is described by phenomenological models. We showed Preisach (a steady-state model) and NLS (transient model) can describe $P$ switching of $HfO_2$ at a low computational cost \cite{Alam_HZO_arbitrary_exciation}. 
According to NLS model the switching time constant $\tau$ of $j^{th}$ hysteron is given by Merz law as-
\begin{equation}\label{Mertz_law}
\tau_j = \tau_0 \exp{ \left( \frac{E_a}{E} \right) ^\alpha }
\end{equation}

where $E_a$ and $E$ are the activation field and the applied electric field, respectively, and $\tau_0$ is the characteristic nucleation time and $\alpha$ is a phenomenological constant. The time dependent polarization is then given by-

\begin{equation}\label{P_from_NLS}
P(t) = 1- \langle \exp{\bigg\{- \left( \frac{t}{\tau_j}  \right)^n \bigg\}} \rangle 	
\end{equation}

%In the following section we show how such phenomenological models can be connected with the double-well potential obtained from the NEB simulation using transition-state theory that does neither predict the S-curve nor QSNC. 

In the following section we show such phenomenological models come out from the double-well potential when used together with transition-state theory. In addition, that does neither predict the S-curve nor QSNC.

% ==== FIG 3

% === III. Schottky-Diode Class-C Rectifier =======================================
% =================================================================================
%\section{Transition-state-theory using Master equation}
\section{Transition-state-theory}

\begin{figure}
  \begin{center}
  \includegraphics[width=3.4in]{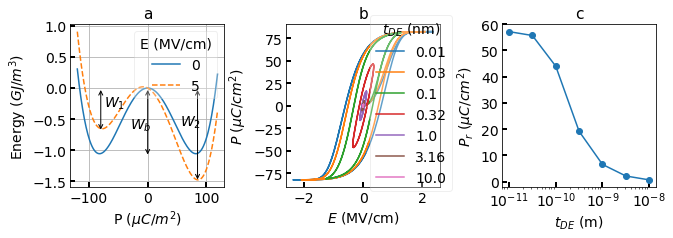}
 % \vspace{-15pt}
  \caption{a) Double well energy profile under applied electric field. Solution of master equation in FE-DE stack gives b) $P-E$ behavior of FE, c) $P_r$ as a function of dielectric thickness }\label{Master_eq_solution_FEDE}
  \end{center}
\end{figure}

The minima of the Landau potential profile (Fig. \ref{Landau_profile_from_NEB}b) lead to the minimum-energy atomic arrangements. Therefore, the minima denote a thermodynamic state where a hysteron can be found. All the hysterons in a FE get distributed in the two states. Application of external electric field tilts the energy profile and thereby changes the barrier height between the states as shown in Fig. \ref{Master_eq_solution_FEDE}a. From here the transition state theory starts with the following postulates-

\begin{itemize}
  \item Switching between the states is an Arrhenius process.
  \item Barriers $W_1$ and $W_2$ are “kinetic barriers” for switching.  
%  \item A hysteron cannot be stabilized on top of the barrier. Barrier determines the transition rate.
\end{itemize}

Let us call the two states as state-1 and state-2 respectively. If all the hysterons were poled to state-1 or state-2 then the net polarization of the system would be −$P_r$ or +$P_r$ respectively. Merz law (eq. \ref{Mertz_law}) has been shown to be equivalent to the Arrhenius process of crossing a barrier (i.e., transition-state-theory) \cite{Depolarization_Multidomain_FE}. 

$W_{1,2}$ as a function of applied electric field is given by-
\begin{equation}\label{Barrier_height_under_E_field}
\begin{split}
W_1 = V(W_b-mqE/A) \\ 
W_2 = V(W_b+mqE/A)
\end{split}
\end{equation}

Here $A$ and $V$ denotes the area and volume of the hysteron respectively, $W_b$ is the barrier height at zero field, $q$ is the elementary charge, $m$ is a number giving the amount of polarization charge (in the unit of $q$) that a hysteron contains. 
The time evolution of the occupation probability of states in a multi-state system is given by Pauli's master equation \cite{Thermally_Activated_switching_kinetics} as-

%\begin{equation}\label{Master_equation}
%\begin{split}
%\frac{dp_1}{dt} = a_{1 \leftarrow 2} p_1 - a_{2 \leftarrow 1} p_2 = \nu_0 \exp{\frac{-W_2}{kT}} p_1 - \nu_0 \exp{\frac{-W_1}{kT}} p_2  \\ 
%\frac{dp_1}{dt} = -a_{1 \leftarrow 2} p_1 + a_{2 \leftarrow 1} p_2= -\nu_0 \exp{\left( \frac{-W_2}{kT} \right)} p_1 + \nu_0 \exp{\left( \frac{-W_1}{kT} \right)} p_2 
%\end{split}
%\end{equation}

\begin{equation}\label{Master_equation}
\begin{split}
\frac{dp_1}{dt} = a_{1 \leftarrow 2} p_1 - a_{2 \leftarrow 1} p_2 ;~  a_{1 \leftarrow 2}=\nu_0 \exp{\left( \frac{-W_2}{kT} \right)} \\ 
\frac{dp_1}{dt} = -a_{1 \leftarrow 2} p_1 + a_{2 \leftarrow 1} p_2 ;~ a_{2 \leftarrow 1}=\nu_0 \exp{\left( \frac{-W_1}{kT} \right)}
\end{split}
\end{equation}

Here $a_{l \leftarrow m}$ is the transition rate from state $m$ to $l$, $\nu_0$ is escape frequency, $k$ is the Boltzmann constant and $T$ is the temperature. From the solution of eq. \ref{Master_equation}, total polarization of the FE is obtained as-
\begin{equation}\label{P_from_Master_equation_solution}
    P=-P_rp_1 + P_rp_2
\end{equation}

%  \caption{Source-pull contours with available input power to the diode set to 6\,dBm.  The impedance is referenced to the junction capacitance of the diode, therefore the lead inductance of the package has been compensated for. Setting $R_{DC}$ to 1080\,$\Omega$ was found to result in the optimal efficiency for this input power.}\label{lpcontours}

At a constant electric field an analytical solution of the eq. \ref{Master_equation} is obtained as-
\begin{equation}\label{Sol_Constant_field}
\begin{split}
p_1(t)=p_1(0)e^{\left(\frac{t}{\tau} \right)} + p_1(\infty) \left(1- e^{\left(\frac{t}{\tau} \right)} \right) \\
p_2(t)=1-p_1(t)
\end{split}
\end{equation}

Here $p_i(0)$ and $p_i(\infty)$ are the fractional occupation of state-$i$ at $t=0$ (initial) and $t=\infty$ (at steady-state) respectively, and $\tau=(a_{1\leftarrow 2}+a_{2\leftarrow 1})^{-1}$ is a time constant for switching. 
The Preisach model assigns a distribution of the coercive field to $p_1(\infty)$ and $p_2(\infty)$ for many hysterons. The remaining term in eq. \ref{Sol_Constant_field} is $\left(1- e^{\left(\frac{t}{\tau} \right)} \right)$. Comparing it against eq. \ref{P_from_NLS} it is seen that the NLS model uses the very same equation with an assumption that $\tau$ is distributed over an exponentially broad spectrum. Overall response of the system is obtained by summing-up the contribution of all the $\tau$ values.
In other words, equation \ref{Sol_Constant_field} encompasses both the Preisach and the NLS models within itself, provided that statistical distributions are introduced. This way it can bridge the physics-based NEB model with the statistical models (that have accurate predictability of experimental results) while automatically absorbing Landau double-well potential in it. Next critical question to answer is, whether such interpretation predicts the existence of QSNC under CBC in a FE-DE stack.

% === IV. Transistor Class-F inv Rectifier ========================================
% =================================================================================
\section{Ferroelectric-Dielectric stack}

\begin{figure}
  \begin{center}
  \includegraphics[width=3 in]{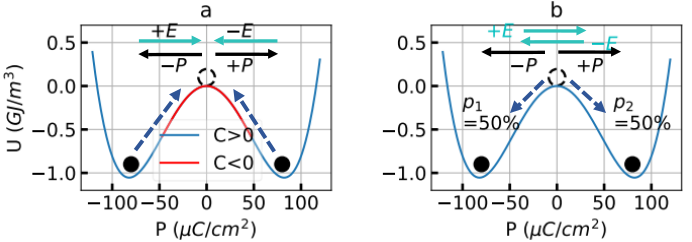}
 % \vspace{-15pt}
  \caption{Difference in the interpretation of zero net-polarization state of ferroelectric-dielectric stack caused by the depolarization field $E$ (eq. \ref{E_depolarization})- a) according to QSNC theory, $E$ pushes FE from the minima to the TOB b) according to transition-state theory, $E$ equally distributes the hysterons in both wells.}\label{Interpretation}
  \end{center}
\end{figure}

The electric field experienced by the FE in a FE-DE stack is given by
\begin{equation}\label{E_depolarization}
    E=\frac{\epsilon_0\epsilon_{DE}}{t_{DE}}\frac{V_{stack}-P}{\epsilon_0 \left(\epsilon_{FE}+\epsilon_{DE}\frac{t_{FE}}{t_{DE}} \right)}
\end{equation}
Here, $V_{stack}$ is the voltage applied across the ferroelectric-dielectric stack, $P$ is the polarization given by equation \ref{Master_equation} and \ref{P_from_Master_equation_solution}, $t_{FE}$,$t_{DE}$ and $\epsilon_{FE}$, $\epsilon_{DE}$ are thicknesses and relative permittivity of FE and DE layers respectively. Therefore, equations \ref{Barrier_height_under_E_field}, \ref{Master_equation}, \ref{P_from_Master_equation_solution} and \ref{E_depolarization} are coupled. We solve the coupled equation for sinusoidal $V_{stack}$. Fig. \ref{Master_eq_solution_FEDE}b shows the $P$ as a function of field in the FE for $t_{FE}=20nm$ and various values of $t_{DE}$. Note the  behavior obtained this way (Fig. \ref{Master_eq_solution_FEDE}b) is always hysteretic and does not show an S-curve (Fig. \ref{Landau_profile_from_NEB}c). Consequently, it does not predict the existence of QSNC. It is because in the Master equation approach, only the minimum in the potential profile represents a physical state where a hysteron can be found.
The polarization at zero electric field,  is extracted from Fig.  \ref{Master_eq_solution_FEDE}b and plotted in Fig. \ref{Master_eq_solution_FEDE}c. When the thickness of the DE increases the polarization decreases with it. It is because thicker dielectric increases the depolarization field that suppresses the polarization. Note that this is the simplest case with an ideal FE-DE interface. In reality, strain boundary condition, interface quality etc. will also play a role in determining the polarization state. The trend has a good qualitative agreement with experimental results reported in \cite{Depolarization_Multidomain_FE}. When the dielectric becomes too thick, the polarization of the ferroelectric becomes zero. 

Note that the same is concluded/predicted by the QSNC theory as well. However, the underlying philosophy is completely different. The comparison is illustrated in Fig. \ref{Interpretation}.
The QSNC theory assumes the maximum in the Landau energy landscape is a physically accessible as well as stabilizable state. The depolarization field that appears in a FE-DE stack pushes the FE towards that maximum (Fig. \ref{Interpretation}a) and keeps it stable on top of the barrier. In the physical sense, when the centrosymmetric position is stabilized, it would correspond to a single-well potential for a particle vibrating around the centrosymmetric  position and not by a double-well. This single-well is describing the dynamics of a (high-k) dielectric system and not a ferroelectric/double-well system: the flatter the single-potential well, the lower the vibrational frequency, therefore higher the dielectric response of the material. In essence, referring to the QSNC as a misnomer for the high-k dielectric suggests an oversimplified interpretation that doesn't fully capture the complexities of the material behavior.

In the transition-state theory on the other hand, the depolarization field pushes the hysterons towards both minima such that they have a 50-50 occupation (Fig. \ref{Interpretation}b). Equal occupation of states gives a net zero polarization according to equation \ref{P_from_Master_equation_solution}, though switching of individual hysteron is by definition hysteretic \cite{Jan_SISC}. In addition, the barrier does not define the "capacitance", instead it  defines the likelihood of jumping of the hysterons from one state to another.

%One can wonder how to explain capacitance amplification in some FE-DE stack without QSNC. Note that it is already explained elsewhere \cite{Kittl_FEDE_without_NC}, and beyond the scope of this letter.

\section{Conclusion}
Landau phenomenological model for phase transition is widely accepted. A particular way of interpreting the model predicts the existence of quasi-static negative capacitance (QSNC). Specifically, the assumption that the barrier region in between two minima of the Landau potential is stabilizable thermodynamic state leads to such prediction. However Landau model is not a singular model for describing ferroelectricity while other available mathematical models do not predict the existence of QSNC. 
We showed the Landau model originates from first principles and when interpreted in terms of transition-state-theory, unifies different available models of ferroelectrics. Doing so also shows QSNC is not intrinsic to the Landau model rather it depends on the interpretation.

%\section*{Acknowledgment}

%Dr. Reveryrand would like to acknowledge the funding by XLIM, Limoges, France. 
%The authors would like to thank Dr. David Root and Dr. Jean-Pierre Teyssier at Agilent Technologies for the loan of the time-domain nonlinear measurement equipment and TriQuint Semiconductor for the donation of the transistors. 

% or


% ============================================
%\appendices
%Appendix one text goes here %\cite{Roberg2010}.

%\section{}

%\section*{Acknowledgment}

%The authors would like to thank D. Root for the loan of the SWAP. The SWAP that can ONLY be usefull in Boulder...

\ifCLASSOPTIONcaptionsoff
  \newpage
\fi



%\begin{thebibliography}{1}


\bibliographystyle{IEEEtran}
\bibliography{IEEEabrv,Bibliography}

\vfill


\end{document}